\documentstyle[preprint,tighten,aps]{revtex}

\renewcommand{\today}{December 8, 1998}
\newcommand {\be}{\begin{equation}}
\newcommand {\ee}{\end{equation}}
\newcommand {\bea}{\begin{eqnarray}}
\newcommand {\eea}{\end{eqnarray}}
\newcommand {\lam}{\lambda}
\newcommand {\del}{\delta}
\newcommand {\eps}{\epsilon}
\newcommand {\veps}{\varepsilon}
\newcommand {\nn}{\nonumber}
\newcommand {\bmath}[1]{\mbox{\boldmath $#1$}}
\newcommand {\bc}{${\rm B_c}$}
\newcommand {\khs}{\hat{k}^2}

\begin{document}
\preprint{\begin{minipage}{2in}\begin{flushright}
  TRI-PP-98-39 \end{flushright}
  \end{minipage}}
\input epsf
\draft
\title{Mesonic decay constants in lattice NRQCD}
\author{B.D.~Jones and R.M.~Woloshyn}
\address{TRIUMF, 4004 Wesbrook Mall,
  Vancouver, British Columbia, Canada, V6T 2A3}
\date{\today}
\maketitle

\begin{abstract}

Lattice NRQCD with leading finite lattice spacing errors removed is used to
calculate decay constants of mesons made up of heavy quarks. 
Quenched simulations are done with  a tadpole improved gauge action containing
\mbox{plaquette} and six-link rectangular terms. The tadpole factor is estimated 
using the Landau link.
For each of the three values of the coupling constant considered, 
quarkonia are calculated for five masses
spanning the range from charmonium through bottomonium, and 
one set of quark masses is tuned to the \bc .
``Perturbative" and nonperturbative meson masses are compared. One-loop
perturbative matching of lattice NRQCD with continuum QCD for the
heavy-heavy vector and axial vector currents is performed.
The data are consistent with $a f_V \propto \sqrt{M_V\, a}$ 
and $f_{B_c}=420(13)$ MeV.

\end{abstract}
%
%
\section{Introduction}
\label{sec:intro}

The gross features of
quarkonia are well described by quenched lattice NRQCD \cite{1,2,3}. 
However, spin splittings between vector and pseudoscalar mesons
tend to be underestimated---especially
when relativistic corrections are 
included \cite{trot,trotshak,lat98klas}.
Using the quark model,  spin splittings are due to a short-ranged Fermi-Breit
interaction and essentially measure the square of the meson's 
wave function at the origin.
Since they are not well understood within lattice NRQCD it is of interest to
investigate physical quantities which probe the same physics. This is provided
by the mesonic decay constants which, in the quark model, are proportional
to the wave function at the origin.
 
In this paper, the focus is on the vector decay constants of quarkonia and the 
pseudoscalar decay constant of the \bc\
meson. The calculated quarkonia constants can of course be compared with 
experiment---for bottomonium this provides a test of the  wave function at the origin, which
is not possible from the spin splitting since the $\eta_b$ has not been
observed yet.
On the other hand, the \bc\ has only recently been discovered \cite{fermilab}
and its decay constant is not measured;
however,  comparison between our results and previous lattice NRQCD and other
model predictions can be made.

A previous study \cite{davthack} of the vector decay constant found large corrections 
from the perturbative matching between the
continuum and lattice currents. Another study \cite{bodwin}
reported a rather imprecise value
for simulations which included the order $v^2$ relativistic corrections of the vector current.
Previous studies \cite{davies96,kim} of the \bc\ decay constant reported results which did not 
include matching or relativistic corrections in the current itself.
In this paper 
we remove the leading finite 
lattice spacing 
errors in the fermion and gauge actions in a symmetric
 fashion and perform (1)
    more precise simulations with the inclusion of the 
 order $v^2$ relativistically corrected currents and (2) one-loop 
 matching between the lattice and continuum currents at lowest order in $v^2$.

The paper is arranged as follows. In Sec.~\ref{sec:lNRQCD} lattice NRQCD 
is introduced and
tadpole improvement is discussed. Continuum and lattice matrix elements are matched up via perturbation
theory in this paper. Thus, in Sec.~\ref{sec:plNRQCD} lattice perturbation theory is set up. 
Appendix~\ref{sec:appendixA} lists the free gluon propagator and
 the integrand of the amputated axial vector correction. A discussion of the matching with continuum
 QCD is presented in Sec.~\ref{sec:match} (Appendix~\ref{sec:appendixB} shows
 the origin of the `$1/v$' terms in the continuum QCD result). 
 A general discussion of lattice decay constants, along with our conventions,  is
 presented in Sec.~\ref{sec:decay}, and finally our results are discussed in Sec.~\ref{sec:results}.

\section{lattice NRQCD}
\label{sec:lNRQCD}

The fermion NRQCD Lagrangian is discretized in a symmetric fashion with the 
leading finite lattice spacing
errors in the spatial and temporal derivatives removed. All links are tadpole improved by dividing by
$u_0$, the average link in the Landau gauge. The choice of
`Landau link' over `average plaquette' tadpole is made based on spin splitting studies
of quarkonia \cite{trotshak}
where improved scaling behavior was observed.
Tadpole improvement in general improves the matching between perturbation theory
and lattice simulations \cite{viability} (see Table~\ref{tab:u0} for our evidence for this).
The fermion Lagrangian we use is

\begin{mathletters}
\label{lf}
\be
a {\cal L}_F =\psi_t^\dagger \psi_t - 
 \psi_t^\dagger \left(1-\frac{a \delta H}{2}\right)_t
\left(1-\frac{a  H_0}{2 n}\right)_t^n 
\frac{U_{4,t-1}^\dag }{u_0}
\left(1-\frac{a  H_0}{2 n}\right)_{t-1}^n \left(1-\frac{a \delta H}{2}\right)_{t-1} 
\psi_{t-1}\;,
\ee

\be
H_0 = -\frac{\Delta^{(2)}}{2 m}\;,\;\delta H = \frac{a^2 \Delta^{(4)}}{24 m} - 
\frac{a \left ( \Delta^{(2)}
\right)^2}{16 n m^2}
\,,
\ee

\be
u_0=\left\langle \frac{1}{3} Re Tr U_\mu \right\rangle\;,~~~~\partial_\mu A_\mu = 0
\,.
\ee
\end{mathletters}

\noindent
$n$ is the stability parameter  chosen to satisfy $n > 3/(m a)$. $\Delta^{(2)}$ is the
gauge-covariant lattice Laplacian, and $\Delta^{(4)}$ is the gauge-covariant lattice quartic
operator ($\sum_i D_i^4$).\footnote{For the standard definitions of these lattice
derivatives see for
example Ref.\ \cite{trot}.}

The above form of the fermion Lagrangian
is convenient for defining Feynman rules.
In lattice simulations the action leads to an evolution equation for the quark
propagator of the form

\bea
G({\bf x},t)-\left(1-\frac{a \delta H}{2}\right)_t
\left(1-\frac{a  H_0}{2 n}\right)_t^n 
\frac{U_{4,t-1}^\dag }{u_0}
\left(1-\frac{a  H_0}{2 n}\right)_{t-1}^n \left(1-\frac{a \delta H}{2}\right)_{t-1}
G({\bf x},t-1)&&\nn\\
&&\hspace{-10ex}=a \delta^4(x)
\;,\label{eq:qprop}
\eea
where in this equation 
the source has been placed at the origin for convenience and \mbox{$G({\bf x},t)=0$}
 for \mbox{$t<0$}.

The gauge action is tadpole improved with
leading finite lattice spacing errors removed by six-link rectangles \cite{lw1}:

\be
S_G = \beta \sum_{{\rm pl}} \frac{1}{3} {\rm Re} \,{\rm Tr} \left(1-U_{{\rm pl}}\right)
-\frac{\beta}{20 u_0^2}
\sum_{{\rm rt}} \frac{1}{3} {\rm Re} \,{\rm Tr} \left(1-U_{{\rm rt}}\right) \,.
\label{glue}
\ee

 In lattice NRQCD meson masses can not be calculated directly from  
meson time correlation functions. 
However, using meson correlators at zero and nonzero momentum, a
`kinetic' mass is obtained by

\be
E({\bf p})-E({\bf 0})=\frac{{{\bf p}}^2}{2 M_{kin}}\,,
\ee
where $E({\bf p})$ is the simulation energy and the momentum 
${\bf p}$ is expressed in units of $2 \pi/(N a)$, where `$N\,a$' is the spatial extent of the lattice.
\pagebreak
\section{perturbative lattice NRQCD}
\label{sec:plNRQCD}

The Feynman rules are derived from the actions of Eqs.~(\ref{lf}) and (\ref{glue})
by making the replacement 
\be
U_\mu(x)\rightarrow \exp[i a g A_\mu^a(x)T^a]
\label{eq:pident}
\;,
\ee
 and expanding
in $g$. This is not quite true because a gauge fixing term must be 
added to the  gauge action before expanding in $g$. The standard covariant gauge-fixing term in lattice
perturbation theory is
\be
S_{GF}=\frac{1}{2\xi}\sum_x\left[\sum_\mu\left\{
A_\mu^a(x)-A_\mu^a(x-\mu)\right\}\right]^2
\;,
\ee
where the gluon field is to be expanded about the midpoint of the link:
\be
A_\mu^a(x)=\int_{-\pi}^\pi \frac{d^4 q}{(2\pi)^4} A_\mu^a(q) \exp\left[i q\cdot(x+\mu/2)\right]
\;.
\ee
To be complete, the fermion field in momentum space is defined by
\be
\psi(x)=\int_{-\pi}^\pi \frac{d^4 p}{(2\pi)^4}\psi(p)\exp\left(i p\cdot x\right)
\;.
\label{complete}
\ee
The gluon propagator follows from the quadratic part of $S_G+S_{GF}$. 
See  Appendix~\ref{sec:appendixA}
for its explicit form.
Note that we use the Feynman gauge ($\xi=1$) for perturbative lattice NRQCD self-energy and
vertex corrections in this paper.
 
The coupling $g$ in this
perturbative identification [Eq.\ (\ref{eq:pident})] that we use is the so called
 ``boosted coupling." It is defined by rewriting the gauge action as
\be
S_G = \frac{6}{g^2}\left[ \frac{5}{3 }\sum_{{\rm pl}} \frac{1}{3} {\rm Re} \,{\rm Tr} 
\left(1-\frac{U_{{\rm pl}}}{u_0^4}\right)
-\frac{1}{12 }
\sum_{{\rm rt}} \frac{1}{3} {\rm Re} \,{\rm Tr} \left(1-\frac{U_{{\rm rt}}}{u_0^6}\right) \right]
\label{glue2}\;.
\ee
The $5/3$ and $-1/12$ are necessary to match the continuum action with its usual
normalization.
Comparing Eqs.\ (\ref{glue}) and (\ref{glue2}) gives
\be
\alpha=\frac{g^2}{4\pi}=\frac{5}{3}\left(\frac{6}{\beta \,u_0^4\, 4\pi}\right)
\;.\label{eq:bcoup}
\ee
The benefits of using this boosted coupling can be seen by comparing the Landau link
calculated nonperturbatively and perturbatively \cite{viability}: Define
$u_0^{(2)}$ by
\be
u_0=1-\alpha \,u_0^{(2)}
\;,
\ee
where $u_0$ is calculated nonperturbatively and $\alpha(\beta,u_0)$ is given by
Eq.\ (\ref{eq:bcoup}). Then compare with a perturbatively calculated
$u_0^{(2)}$. This comparison is shown in Table~\ref{tab:u0}; the close agreement
justifies the use of a boosted coupling.

Given the Feynman rules, the perturbative matching factors for the decay constants
follow. We discuss this matching in detail in the next section.
A ``perturbative" meson  mass can also be defined. For example in the equal mass case
\be
M_{pert}=2 (m Z_m-E_0)+E_{sim}
\;,
\ee
where $Z_m$ is the mass renormalization and $E_0$ is the energy shift of a quark,
and $E_{sim}$ is the simulation energy extracted from the meson correlator.
For details on how $Z_m$ and $E_0$ are defined see Ref.\ \cite{morning1}.

\section{Matching with continuum QCD}
\label{sec:match}

Lattice NRQCD is an effective field theory that is fundamentally
different from continuum QCD---the ultraviolet divergences 
in the matrix elements of interest are not the same. 
After renormalization, these differences are finite (since the
infrared divergences are the same), but
nevertheless in order to obtain continuum QCD results,
a matching step must be performed. In this section we will focus
on the \emph{axial} vector current matching since the vector current case follows
in a similar manner. Details of the axial vector current matching
will be given, and in the end, results for the vector current case will 
also be shown.

In general for the annihilation
decay of the \bc\ meson, matching leads to
 \be
 \left.\langle 0| \overline{b} \gamma_0 \gamma_5 
 c|c \overline{b}\rangle\right |_{cont}=Z_{match}\left.\langle 0| \overline{b}
  \gamma_0 \gamma_5 
 c|c \overline{b}\rangle\right |_{lat}+{\cal O}(v^2)\;,
 \ee
where $v$ is the relative velocity of the $\overline{b}$ and $c$.
In performing this matching in this paper we neglect the order $v^2$ terms and calculate $Z_{match}$ in
one-loop perturbation theory. For this system $\alpha \sim .25$ and
$v^2 \sim .2$: the expansion parameters are somewhat less than unity.

Although the matching can be done in one step,\footnote{We performed the 
matching in one step also (matching continuum QCD directly with lattice NRQCD) 
and obtained the same result as what follows from the discussion below.} 
we will use two steps for pedagogical reasons. 
The two step matching procedure starts with the matching of 
lattice NRQCD (lNRQCD)  with
continuum NRQCD (cNRQCD). Then continuum NRQCD is matched with continuum QCD (cQCD).
In the end a one-loop formula, relating the simulated lattice NRQCD matrix element
with its continuum counterpart of full QCD, is obtained. 

We will use the following notation:
the amputated vertex correction will be written as `$g^2\delta V$';
including the wave function renormalization factors, the total matrix element will
be written as
 `$1+g^2\left(\delta V+\delta Z_b/2+\delta Z_c/2\right)$' times the tree level amplitude;
 and we will use different regulators as given below---but our renormalization
 scheme is always the on-mass-shell scheme.


\subsection{Matching lattice NRQCD with continuum NRQCD}

The current matrix element is ultraviolet finite in NRQCD. However infrared divergences
do arise, and in this subsection a gluon mass is used as a regulator.
The matrix elements are calculated in the limit
\be
1 \gg \lambda/m_{red} \gg  v \longrightarrow 0\;,
\label{eq:scales}
\ee
where $\lambda$ is the gluon mass, $v$ is the relative velocity  and $m_{red}$ is
the reduced mass of the system,
\be
1/m_{red}=1/m_b +1/m_c
\;.
\ee

In continuum NRQCD it is convenient to work in the Coulomb gauge. The transverse
gluons coupling to the quarks is suppressed by powers of $v$, so only the Coulomb exchange
term needs to be considered. 
The wave function renormalization factor is 
\be
\left.g^2\delta Z\right|_{cNRQCD}=-i C_f g^2 \int \frac{d^4k}{(2\pi)^4}
\frac{1}{({\bf k}^2+\lam^2)(k_0+\frac{{\bf k}^2}{2 m}-i \veps)^2}\;,\label{zzzzz}
\ee
where $C_f=4/3$.
Performing the $k_0$ integration, we see that the result is trivial, $\left.\delta Z\right|_{cNRQCD}=0$.
The amputated vertex correction is
\be
\left.g^2\del V\right|_{cNRQCD}=i C_f g^2 \int \frac{d^4k}{(2\pi)^4}
\frac{1}{({\bf k}^2+\lam^2)(k_0-\frac{{\bf k}^2}{2 m_b}+i \veps)
(k_0+\frac{{\bf k}^2}{2 m_c}-i \veps)}\;.
\ee
Performing the $k_0$ integration, we are left with 
\be
\left.g^2\del V\right|_{cNRQCD}=2 m_{red} C_f g^2 \int \frac{d^3k}{(2\pi)^3}
\frac{1}{({\bf k}^2+\lam^2){\bf k}^2}=\frac{2  m_{red} \,g^2}{3\pi\lam}
 \;.\label{ntnt}
\ee
In summary, for continuum NRQCD with the scales
proportioned as in Eq.~(\ref{eq:scales}) through one loop we have
\be
 \left.\langle 0| \chi_b \psi_c|c \overline{b}\rangle\right |_{cNRQCD}=
 \eta^\dag_b \xi_c\left[
 1+\frac{2  m_{red}\, g^2}{3\pi\lam}
 \right]
\;,
\label{eq:cnrqcd}
\ee
where $\psi_c$ and $\chi_b$ are non-relativistic $c$ and ${\overline b}$ fields respectively.

Lattice NRQCD is defined in Euclidean space.
However,
since the zeroth component of the axial vector current is identical
in Euclidean and Minkowski space, one can match directly with the continuum
NRQCD result.\footnote{For the vector case there is just a trivial factor of $i$ that
must be kept track of: $\bmath{\gamma}_{_{Eucl}}=-i \bmath{\gamma}_{_{Mink}}$.}
The detailed form of the integrands is not given here (see  Appendix~\ref{sec:appendixA}). 
Instead, we will just write the form of the final result
\be
\left.\langle 0| \chi_b \psi_c|c \overline{b}\rangle\right |_{lNRQCD}=
 \eta^\dag_b \xi_c\left[
 1+g^2
 \left(\frac{2 m_{red}}{3\pi\lam}+\del \overline{V}_{lat}+\del {Z}_{lat}^c/2+
 \del{Z}_{lat}^b/2\right)
 \right]
\;,
\label{eq:lnrqcd}
\ee
where the bar on $\delta V_{lat}$ signifies the separation of the linear infrared divergence. 
The integration routine {\footnotesize   VEGAS} \cite{vegas} is used to perform 
the integrations. {\em Only infrared
finite integrands are integrated with\/} {\footnotesize   VEGAS}. We subtract and 
add low three-momentum versions (leaving
the energy variable alone) of 
the original integrand, where the subtraction is chosen to (1) make the original integrand
infrared finite and (2) leave a simpler infrared divergent integrand that can be integrated
analytically (or else repeat the subtraction procedure on this new term also). In an equation
\be
\int_{-\pi}^\pi d^4k\, I(k)=\int_{-\pi}^\pi d^4k \,\left[I(k)-I_{sub}(k)\,\theta(c^2-{\bf k}^2)\right]+
\int_{-\pi}^\pi d^4k \,I_{sub}(k)\,\theta(c^2-{\bf k}^2)
\;,
\ee
where the step function is chosen for convenience in the analytic integration step---`$c$' 
is an arbitrary parameter less than `$\pi$' (for example `$1$' or `$2$').
In this way,
 the linear infrared divergence is analytically separated as we have
written it in Eq.~(\ref{eq:lnrqcd}).

To conclude, divide Eq.~(\ref{eq:cnrqcd}) by Eq.~(\ref{eq:lnrqcd}). Through one loop with the scales
proportioned as in Eq.~(\ref{eq:scales}) 
\be
\left.\langle 0| \chi_b \psi_c|c \overline{b}\rangle\right |_{cNRQCD}=
\left.\langle 0| \chi_b \psi_c|c \overline{b}\rangle\right |_{lNRQCD}
\left[
 1-g^2
 \left(\del \overline{V}_{lat}+\del {Z}_{lat}^c/2+
 \del{Z}_{lat}^b/2\right)
 \right]
 \;,
 \label{eq:match1}
\ee
where the remaining matching factor is infrared finite; however,
 a logarithmic divergence cancels
between the remaining terms.
\subsection{Matching continuum NRQCD with continuum QCD}

First, our conventions for this subsection:
dimensional regularization  regulates the infrared (IR) and
ultraviolet (UV) divergences;\footnote{
 The usual $D=4-2 \epsilon$ identification
is made.} 
we will work with finite but small
relative velocity $v$; $\gamma_5$ is chosen to anticommute with all other 
gamma matrices; and as already mentioned,
 our renormalization scheme is
the on-mass-shell scheme.

The necessary matching calculation for this subsection 
has been performed already by Braaten and Fleming \cite{braaten}.
Details are omitted here and can be found in Ref. \cite{braaten}.
However, one intermediate step of the calculation is discussed in 
Appendix~\ref{sec:appendixB}, that is, the origin of the `$1/v$' terms in 
the results. 

The one-loop renormalization of the axial vector current matrix element 
consists of amputated vertex  and
 wave function renormalization corrections. In continuum NRQCD it is convenient to use the
 Coulomb gauge. Here the wave function renormalization correction
vanishes [see Eq.~(\ref{zzzzz})], and the amputated vertex 
correction is UV finite. However there is an IR divergence.
The full result in continuum NRQCD is
\begin{mathletters}
\label{eq:conttot}
\bea
\left.\langle 0| \chi_b \psi_c|c \overline{b}\rangle\right |_{cNRQCD}&=&
\eta^\dag_b \xi_c\left[
1+\frac{g^2}{6\pi^2} \left\{
\frac{\pi^2}{v}
\right.\right.\nn\\
&&\left.\left.\hspace{5ex}\mbox{}
-\frac{i\pi}{v}\left(
\frac{1}{\epsilon_{IR}}-2\,\log\frac{2 m_{red} v}{\mu}+\log4\pi
-\gamma
\right)
\right\}
\right]\;.\label{e1}
\eea

In continuum QCD it is convenient to use the Feynman gauge.
The matrix element is UV divergent, but these divergences cancel between
the amputated vertex and wave function renormalization factors. There is a real
IR divergence that also cancels, but the final result has an imaginary IR divergence.
The full result in continuum QCD is
\bea
\left.\langle 0| \overline{b} \gamma_0 \gamma_5 
 c|c \overline{b}\rangle\right |_{cQCD}&=&\overline{v}_b \gamma_0 \gamma_5 
 u_c \left[
1+\frac{g^2}{6\pi^2} \left\{
\frac{\pi^2}{v}+\frac{3}{2}\frac{m_b-m_c}{m_b+m_c} \log\frac{m_b}{m_c}-3
\right.\right.\nn\\
&&\left.\left.\hspace{5ex}\mbox{}
-\frac{i\pi}{v}\left(
\frac{1}{\epsilon_{IR}}-2\,\log\frac{2 m_{red} v}{\mu}+\log4\pi
-\gamma
\right)
\right\}
\right]
\;.\label{e2}
\eea
\end{mathletters}

For the respective one-loop integrals to converge 
$\epsilon_{IR}<0$ and $\epsilon_{UV}>0$.
Also, the diagrams are evaluated slightly above
threshold, 
\be
q^2=(m_b+m_c)^2+m_b m_c v^2+{\cal O}(v^4)\;,
\ee
thus the $\overline{b}$ and $c$ can simultaneously go on-mass-shell giving rise to
 the imaginary contributions.
Dividing these continuum QCD and NRQCD results [Eq.~(\ref{eq:conttot})], 
through one loop and for small $v$,
gives for the final result
\be
\left.\langle 0| \overline{b} \gamma_0 \gamma_5 
 c|c \overline{b}\rangle\right |_{cQCD}=
 \left.\langle 0| \chi_b \psi_c|c \overline{b}\rangle\right |_{cNRQCD}
 \left[
1+\frac{g^2}{6\pi^2} \left(
\frac{3}{2}\frac{m_b-m_c}{m_b+m_c} \log\frac{m_b}{m_c}-3
\right)\right]
\;, \label{eq:match2}
\ee
where `$\overline{v}_b \gamma_0\gamma_5 
 u_c=\eta^\dag_b \xi_c+{\cal O}(v^2)$' has been used.
\subsection{Matching lattice NRQCD with continuum QCD}

In both of the previous two subsections  the ultraviolet
and infrared divergences  canceled in the final result, thus
the two steps can be combined.
Substituting Eq.~(\ref{eq:match1}) into Eq.~(\ref{eq:match2}) gives for our final
result
\bea
\left.\langle 0| \overline{b} \gamma_0 \gamma_5 
 c|c \overline{b}\rangle\right |_{cQCD}=
 \left.\langle 0| \chi_b \psi_c|c \overline{b}\rangle\right |_{lNRQCD}
 &&\left[
1+\frac{g^2}{6\pi^2}\left(\raisebox{0ex}[3ex][2ex]{}
\frac{3}{2}\frac{m_b-m_c}{m_b+m_c} \log\frac{m_b}{m_c}-3\right)\right.\nn\\
&&\left.\hspace{.6in}\mbox{}
-g^2\left(\del \overline{V}_{lat}+\del {Z}_{lat}^c/2+
 \del{Z}_{lat}^b/2\right)\raisebox{0ex}[3ex][2ex]{}
\right]
\;, \label{eq:matchfinal}
\eea
which as already mentioned is the same result we obtained by directly
matching lattice NRQCD with continuum QCD in one step with
the scales proportioned as in Eq.~(\ref{eq:scales}).

As promised, the result for the vector case will also be given. 
The procedure is the same as for the axial vector,
and the result is very similar---take
the previous with equal quark masses and `$-3\rightarrow -4$':
\bea
\left.\langle 0| \overline{Q} \bmath{\gamma} 
 Q|Q\overline{Q}\rangle\right |_{cQCD}=
 \left.\langle 0| \chi_Q \bmath{\sigma}\psi_Q|Q \overline{Q}\rangle\right |_{lNRQCD}
 &&\left[
1+\frac{g^2}{6\pi^2} \raisebox{0ex}[3ex][2ex]{}
\left(-4\right)
-g^2\left(\del \overline{V}_{lat}+\del {Z}_{lat}^Q
 \right)\raisebox{0ex}[3ex][2ex]{}
\right]
\;. \label{eq:matchfinal2}
\eea
We will use the following notation to report these matching contributions:
\be
\left.\left\langle 0|J|M\right\rangle \right|_{cQCD}=Z_{\rm match} 
\left.\left\langle 0|J|M\right\rangle \right|_{lNRQCD}
\;.
\label{eq:heymatch}
\ee
Tables~\ref{tab:zmatchquarkonia} and \ref{tab:zmatchbc} show our results.
\section{decay constants}
\label{sec:decay}

A meson propagator is given by

\begin{eqnarray}
G({\bf p},t)&=&\left\langle \sum_{{\bf x}} \exp [ -i {\bf p} \cdot (
{\bf x}-{{\bf x}}_0)] J({\bf x},t) J^\dagger({{\bf x}}_0,t_0)\right\rangle\nn\\
&=&\left\langle \sum_{{\bf x}} \exp [ -i {\bf p} \cdot (
{\bf x}-{{\bf x}}_0)]
\,{\rm Tr}\left[(\Gamma_x G_{x\,x_0})
(G_{x\,x_0}\Gamma_{x_0})^\dagger\right]
\right\rangle\,,
\end{eqnarray}

\noindent
where the trace is over spin and color, and $G_{x\,x_0}$ is the quark propagator 
of Eq.~(\ref{eq:qprop}). $J({\bf x},t)=\chi_x \Gamma_x \psi_x$ is a non-relativistic  current
with $\Gamma_x=\Omega_x \omega_x$, where $\Omega_x$ interpolates the meson of 
interest\footnote{$\Omega_x=I$, $\bmath{\sigma}$,  and $\bmath{\Delta}$ 
for the ${}^1\!S_0$, ${}^3\!S_1$, and ${}^1\!P_1$ states respectively.} 
and $\omega_x$ is a smearing
operator chosen in a gauge-covariant fashion:

\be
\omega_x=[1+\epsilon \Delta^{(2)}(x)]^{n_s}\,.
\ee

\noindent
We set $\epsilon=1/12$ and tune the  smearing parameter $n_s$ to maximize the
overlap with the state of interest. The range 7--30 for $n_s$ was found 
to be sufficient, with
the P-wave requiring about twice as much smearing as the S-wave 
 and the smearing parameter increasing for decreasing
quark mass.

The asymptotic form of this non-relativistic meson propagator is
\be
G({\bf p},t)~~
\begin{picture}(40,6)(0,0)
\put(0,2){\vector(1,0){40}}
\put(0,-4){\mbox{\scriptsize $t-t_0\rightarrow \infty$}}
\end{picture}~~
\left|
\langle 0 | J(0) | {\bf p}\rangle
\right|^2
\exp [-E({\bf p})(t-t_0)]\,.
\ee

\noindent
In continuum Minkowski notation,
this current matrix element for  a pseudoscalar and vector
 at rest is related to the respective decay constant
by
\begin{mathletters}
\label{eq:decayconstants}
\bea
\langle 0| \overline{b} \gamma_0 \gamma_5 
 c|B_c\rangle&=&\frac{i f_{B_c} M_{B_c}}{\sqrt{2 M_{B_c}}}\;,\\
 \langle 0 |\overline{Q}{\bmath \gamma}Q | V\rangle&=&\frac{i f_V M_V\,{\bmath \epsilon}}{\sqrt{2 M_V}}
\;,
\eea
\end{mathletters}
where ${\bmath \epsilon}$ is a polarization vector
and for matching purposes a non-relativistic norm is assumed.

Simulations are performed with  order $v^2$ 
relativistic corrections included in the
currents.  The relativistically corrected
 interpolating operator for a non-relativistic vector meson is given by

\be
\Omega_V^{rel}=\sigma_i+\frac{1}{8 m^2} \left(
\Delta^{(2)\dagger}\sigma_i+\sigma_i\Delta^{(2)}
\right)
-\frac{1}{4m^2}\left(\bmath{\sigma}\cdot{\bmath{\Delta}}^\dagger\right)\sigma_i
\left(\bmath{\sigma}\cdot\bmath{\Delta}\raisebox{0ex}[2ex][1ex]{}\right)\,,
\label{impv}
\ee

\noindent
where $\bmath{\Delta}$ is the  gauge-covariant symmetric lattice derivative
and $\Delta^{(2)}$ is the
gauge-covariant lattice Laplacian. For the \bc\  the relativistically corrected interpolating operator
is
\be
\Omega_{B_c}^{rel}=1+\frac{\Delta^{(2)}}{8 m_{red}^2}
\,.
\label{impbc}
\ee

Our final decay constants are reported using the following notations: Let `src' represent
L (local) or S (smeared). `X' is ${}^1P_1$, PS (pseudoscalar), or V (vector).
Three correlators are of interest:
\bea
&(1)&~~ C^L_X\equiv \left\langle \sum_{{\bf x}} 
 J^L_X({\bf x},t) {J^L_X}^\dagger({{\bf x}}_0,t_0)\right\rangle
 \equiv Z^L_X \exp\left[-E_X^L(t-t_0)\right]\;,\nn\\
&(2)&~~C^S_X\equiv \left\langle \sum_{{\bf x}} 
 J^L_X({\bf x},t) {J^S_X}^\dagger({{\bf x}}_0,t_0)\right\rangle
 \equiv Z^S_X \exp\left[-E_X^S(t-t_0)\right]\;,\nn\\
{\rm and}~~&(3)&~~C^{S,rel}_X\equiv \left\langle \sum_{{\bf x}} 
 J^{L,rel}_X({\bf x},t) {J^S_X}^\dagger({{\bf x}}_0,t_0)\right\rangle
 \equiv Z^{S,rel}_X \exp\left[-E_X^{S,rel}(t-t_0)\right]\;,
\eea
where $J^{L,rel}_X({\bf x},t)=\chi_x \Omega_{X}^{rel} \psi_x$
and the right-hand-side of these equations assumes $t\gg t_0$.
Given this, we report two forms of decay constants:
\bea
&&\frac{f_X M_X}{\sqrt{2 M_X}}=a^{-\frac{3}{2}} Z_{\rm match} \sqrt{Z_X^L}\\
{\rm and}~~&&\frac{f_X^{rel} M_X}{\sqrt{2 M_X}}=a^{-\frac{3}{2}} Z_{\rm match} 
\frac{\sqrt{Z_X^L}\,Z_X^{S,rel}}{Z_X^S}
\;,
\eea
where $f_X^{rel}$ includes 
the order $v^2$ relativistic corrections in the currents and $Z_{\rm match}$ is
the perturbative matching factor of Eq.~(\ref{eq:heymatch}). The spirit of 
these definitions is from Ref.~\cite{hashimotoetal}.
\section{Results and discussion}
\label{sec:results}

Tables~\ref{tab:bc} and \ref{tab:candb}
 show the simulation parameters and compare
$M_{pert}$ and $M_{kin}$.
 The lattice spacing is fixed by the splitting between S- and P-wave states which
is taken to be 458~MeV for all simulations.
For each coupling, five quark masses spanning the range from
charmonium through bottomonium are used, and one set of quark masses is tuned to the
\bc\ which we take to be 6.35~GeV for the spin averaged ground state.

 The data sample includes 1200  quenched gauge field configurations at
 $\beta =$ 7.4 ($10^3\times 16$), and 1600 configurations at $\beta =$ 7.2 and 
7.3 ($8^3\times14$).
A standard Cabbibo-Marinari pseudo heat bath is used to generate the gauge-field configurations.
After 4000 thermalizing sweeps, the number of updates between measurements is 30.
Autocorrelation times for
the correlation functions are checked and satisfy $\tau \alt \frac{1}{2}$.
Multiple sources (N/2) along the spatial diagonal are used to measure the
 local-smeared meson correlators with the number of
smearing steps optimized to have the best overlap
with the state of interest. Single exponential fits to the 
correlation functions are used to get the best
estimates of the masses and decay constants. 
Effective mass plots are generated and used to
choose the interval in which to do the fits. In all cases acceptable Q values
are obtained.
All statistical errors are estimated by the bootstrap method
with twice as many bootstrap ensembles as
there are configurations. 
 As expected, simulation energies from L-L, L-S, and
Lrel-S correlators are all consistent with each other.

\begin{figure}
\begin{center}
\leavevmode
\epsfxsize=6.0in\epsfbox[66  131  513  630]{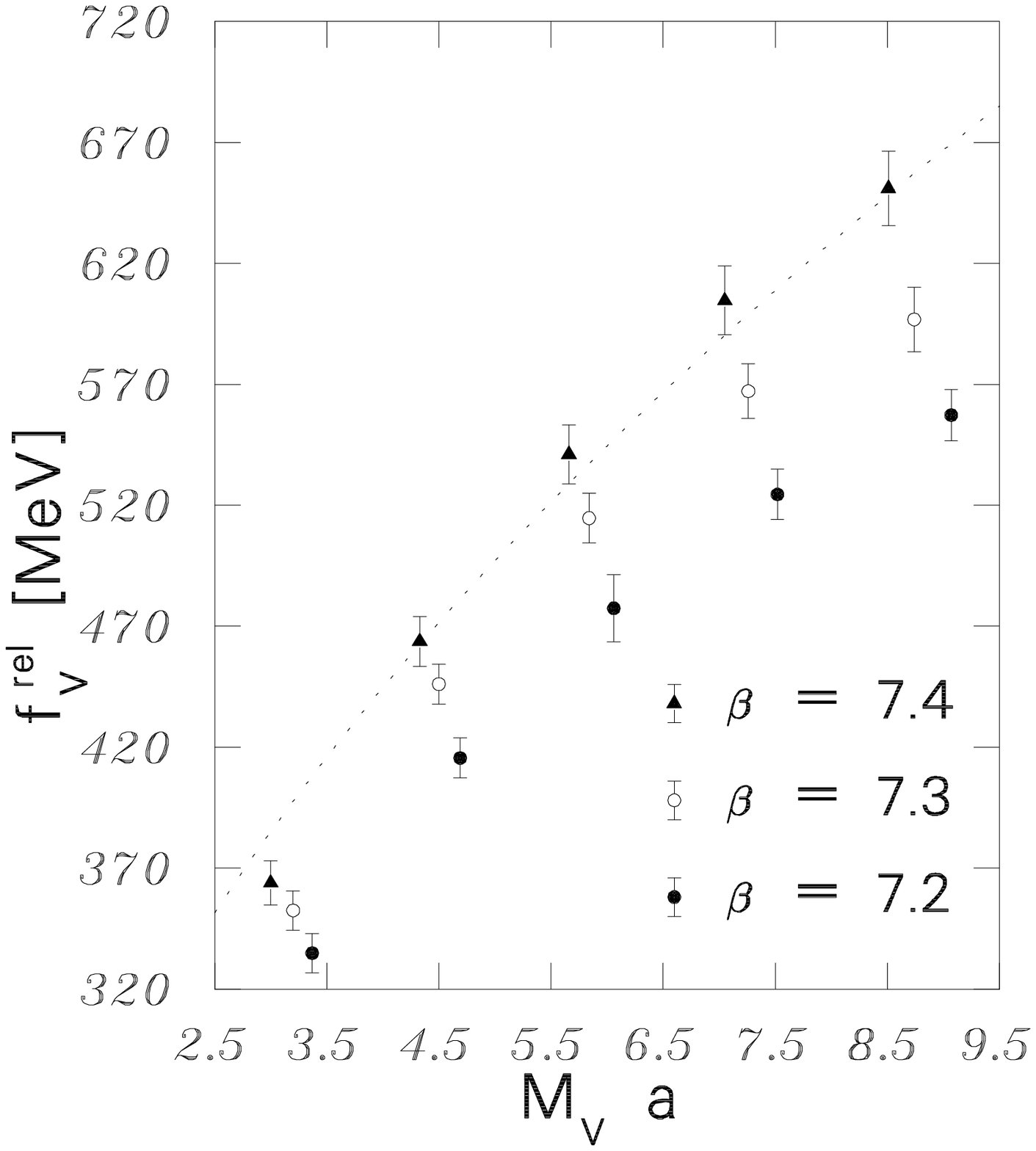}
\end{center}
\caption{
 Meson mass dependence of the order $v^2$ relativistically corrected vector decay constants 
 of quarkonia including
the one-loop perturbative matching. The dotted line is proportional to $\sqrt{M_V \,a}$.
\label{fig:fvsmv}}
\end{figure}

Fig.~\ref{fig:fvsmv} shows our main  result for quarkonia:
\be
a f_V \propto \sqrt{M_V\, a}
\;.\label{main}
\ee
For a tabular form of this information see Table~\ref{tab:quarkoniadecay}.
This square-root dependence is quite interesting.
Shortly after the 
 discovery of charm, Yennie \cite{sakurai} noticed 
 this same dependence from empirical data:
\be
\frac{\Gamma^V_{e{\overline e}}}{e_q^2}\sim {\rm constant} \sim 12 \;{\rm keV}
\ee
for light through heavy ground-state vector mesons ($e_q$ is the quark charge in 
units of $e$). This is the same as our data 
since the leptonic width is proportional to $
e_q^2\frac{f_V^2}{M_V}
$.
 A final note on this  result:
 a \emph{linear}
 (\emph{Coulomb}) potential  implies a 
\emph{constant} (\emph{linear})
 dependence on the meson mass for the decay constant, so our data is consistent with
 a superposition of the two potentials.

\begin{figure}
\begin{center}
\leavevmode
\epsfxsize=6.0in\epsfbox[67  135  513  660]{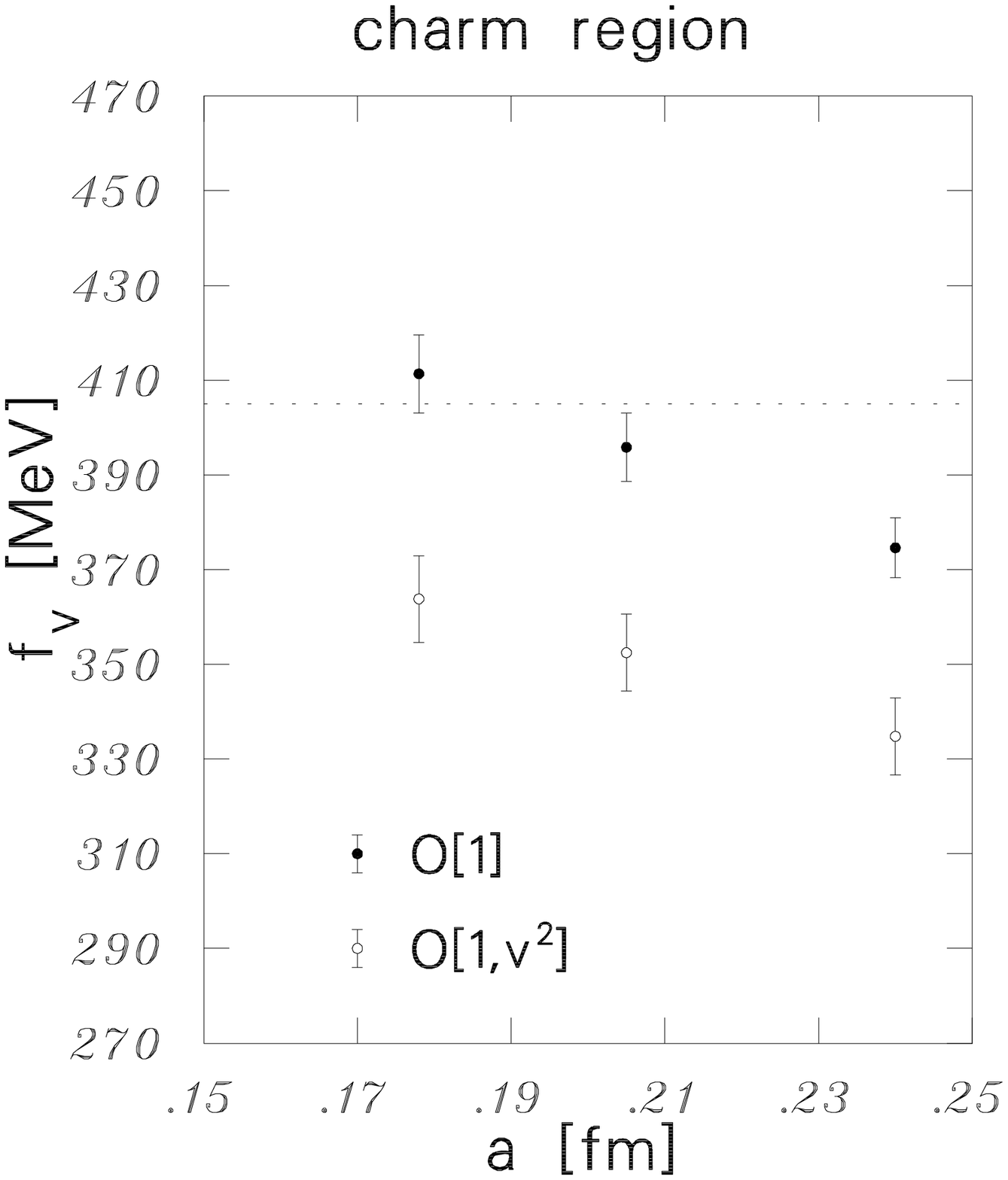}
\end{center}
\caption{
 The calculated vector decay constant of charmonium 
 (including the perturbative matching) at different 
values of the lattice spacing with and without the relativistic corrections in the current.
The dotted line is the experimental result.
\label{fig:fc}}
\end{figure}

\begin{figure}
\begin{center}
\leavevmode
\epsfxsize=6.0in\epsfbox[67  135  512  660]{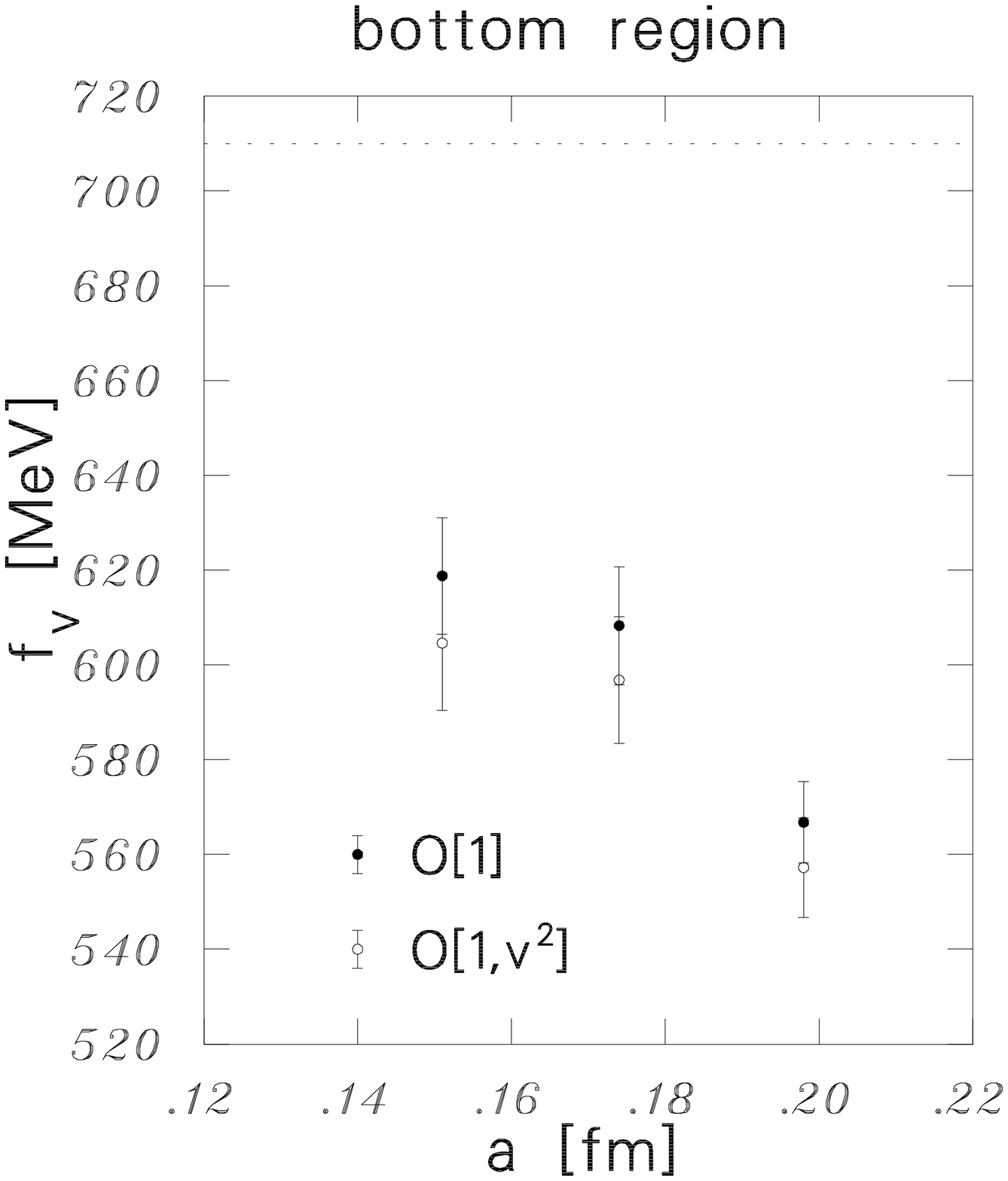}
\end{center}
\caption{
 The calculated vector decay constant of bottomonium 
 (including the perturbative matching) at different 
values of the lattice spacing with and without the relativistic corrections in the current.
The dotted line is the experimental result.
\label{fig:fb}}
\end{figure}

In order to compare with experimental values of quarkonia we use results from 
Fig.~\ref{fig:fvsmv} with masses nearest the physical
$J/\psi$ and $\Upsilon$ masses. However, the masses are not tuned precisely
to the experimental values (see Table~\ref{tab:candb}). For example, the 
$\beta=7.2$ decay constants should be larger according to the derived 
$a f_V \propto \sqrt{M_V\, a}$ dependence.
With this caveat, Figs.~\ref{fig:fc} and \ref{fig:fb} show 
the comparison of our calculated decay constants with empirical values.
Previous work on spin splittings would lead us to expect decay constants smaller
than experimental values and this is what we find although the 10--15\% underestimation
(with the order $v^2$ relativistically corrected currents) may be a bit less than what one might have
anticipated.
As expected---since $v_c^2\sim .3$ and $v_b^2\sim .1$---the
 shift from the order $v^2$ relativistic corrections in the currents
is approximately three times as big for charmonium as compared to bottomonium.

\begin{figure}
\begin{center}
\leavevmode
\epsfxsize=6.0in\epsfbox[67  135  512  660]{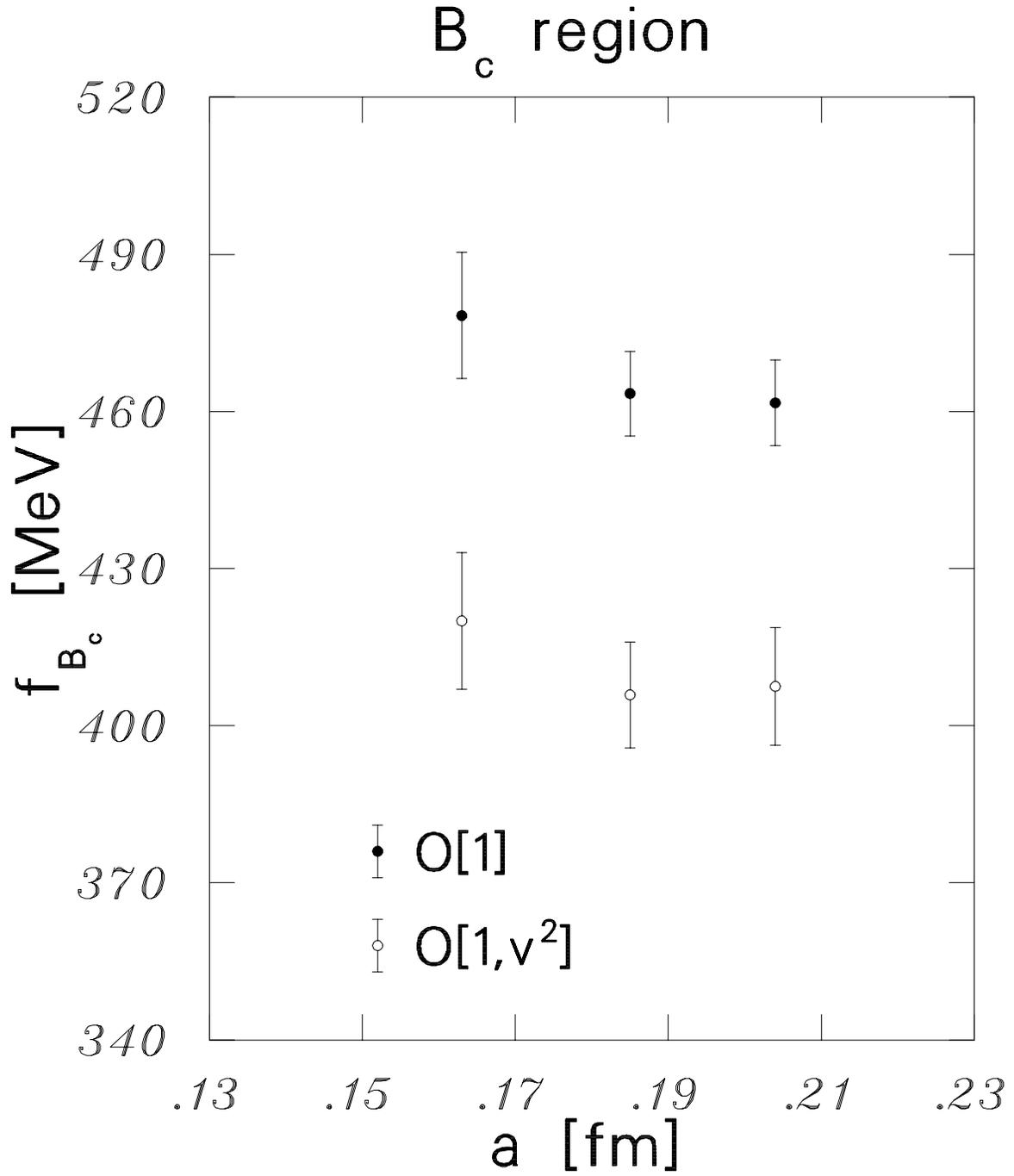}
\end{center}
\caption{
 Scaling behavior of the \bc\
decay constant (including
the perturbative matching) with and without the relativistic corrections in the current.
\label{fig:fbc}}
\end{figure}

\begin{figure}
\begin{center}
\leavevmode
\epsfxsize=6.0in\epsfbox[92   58  567  679]{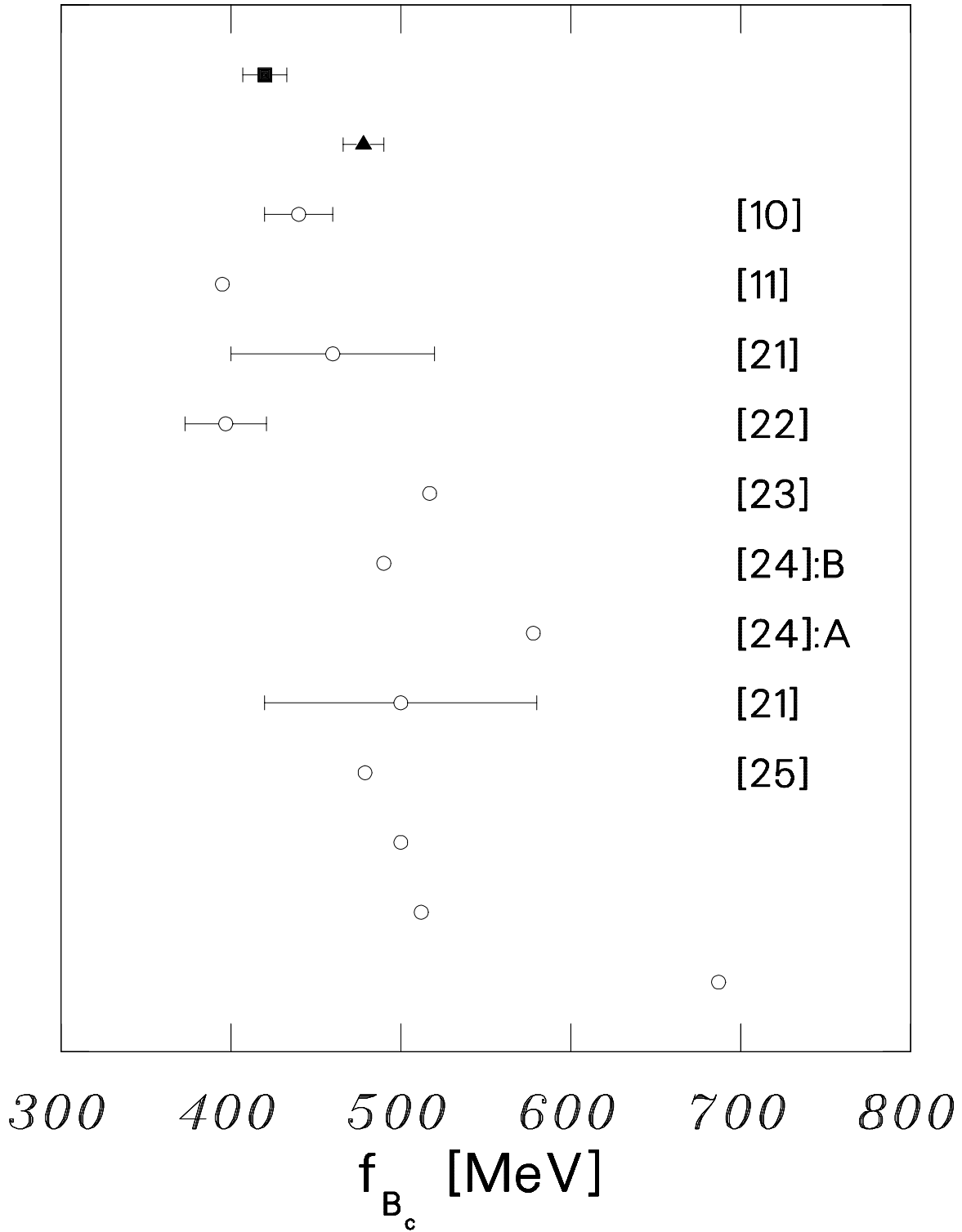}
\end{center}
\caption{
A comparison with previous work. The filled square is
this work: ${\cal O}(1,v^2)$;  filled triangle is this work: ${\cal O}(1)$;
 and the open circles are other models labeled by their respective reference.
\label{fig:fbccompare}}
\end{figure}

  The main \bc\ (spin averaged ground state $\sim$ 6.35~GeV; see Table~\ref{tab:bc}) 
  results  are shown in Fig.~\ref{fig:fbc}.
The same information is in Table~\ref{tab:bcdecay}.
  In Fig.~\ref{fig:fbccompare} 
a comparison with previous lattice results \cite{davies96,kim} for $f_{B_c}$ combined with
a partial list of other model results \cite{gershtein2,gershtein1,fulcher,abdelhady,eichtenquigg} 
is shown. Our final result,
 420(13) MeV, is taken from the $\beta=7.4$ data point ($a\sim.16$ fm). 

This is an initial study of  decay constants made up of heavy quarks 
which systematically includes  removal of the
leading finite lattice spacing errors in the action, and measures the
effects of the leading relativistic corrections in the vector and axial vector currents.
The current matrix elements are matched with continuum QCD through
one-loop in perturbation theory at lowest order in $v^2$. After removal of the
leading finite lattice spacing errors in the action we find these matching
corrections to be small. This is shown in  Tables~\ref{tab:zmatchquarkonia} and \ref{tab:zmatchbc}:
10\% for the $J/\Psi$; 5\% for the $\Upsilon$; and less than
1\% for the \bc .

\acknowledgments

We  thank R. Lewis, N. Shakespeare, and  H. Trottier for useful discussions. 
We also thank H. Trottier for the use of his quark propagator code.
This work is supported by the Natural Sciences and Engineering Research
Council of Canada. 

\pagebreak
\appendix

\section{Gluon propagator and $\delta V_{lat}$ for the ${\rm B_c}$}
\label{sec:appendixA}

The gluon propagator follows from the
quadratic part of $S_G+S_{GF}$. In the Feynman gauge ($\xi=1$) the
gluon propagator is \cite{WeiszWohlert} ($\lambda$ is our gluon mass)
\be
D_{\mu\nu}(k)=D_{\nu\mu}(k)=\frac{1}{\hat{k}^2}\frac{1}{\hat{k}^2+\lambda^2}\left[
\hat{k}_\mu\hat{k}_\nu+\sum_\sigma (\hat{k}_\sigma \delta_{\mu\nu}-\hat{k}_\nu \delta_{\mu\sigma})
\hat{k}_\sigma A_{\sigma\nu}(k)\right]\;,
\ee
with ($c_1=-1/12$)
\bea
A_{\mu\nu}(k)=A_{\nu\mu}(k)&=&(1-\delta_{\mu\nu})\Delta(k)^{-1}\left[
(\khs)^2-c_1 \khs\left(2\sum_\rho\hat{k}_\rho^4 +\khs \sum_{\rho\neq\mu ,\nu}\hat{k}_\rho^2\right)
\right.\nn\\
&&\hspace{5ex}\left.+c_1^2\left\{\left(\sum_\rho\hat{k}_\rho^4\right)^2+\khs\sum_\rho\hat{k}_\rho^4
\sum_{\tau\neq\mu ,\nu}\hat{k}_\tau^2+(\khs)^2\prod_{\rho\neq\mu ,\nu}\hat{k}_\rho^2\right\}\right]
\;,
\eea
where
\bea
\Delta(k)&=&\left(\khs-c_1 \sum_\rho \hat{k}_\rho^4\right)\left[
\khs-c_1\left\{(\khs)^2+\sum_\tau\hat{k}_\tau^4\right\}+\frac{1}{2}c_1^2\left\{
(\khs)^3+2\sum_\tau\hat{k}_\tau^6-\khs\sum_\tau\hat{k}_\tau^4\right\}\right]\nn\\
&&\hspace{10ex}-4 c_1^3\sum_\rho\hat{k}_\rho^4\prod_{\tau\neq\rho}\hat{k}_\tau^2
\;.
\eea
The usual 
\be
\hat{k}_\mu=2 \sin (k_\mu/2)
\ee
and 
\be
\hat{k}^2=4\sum_{\mu=1}^4 \sin^2 (k_\mu/2)
\ee
definitions have been made.
Also note that although we are working in the Feynman gauge here,
$D_{\mu\nu}$ is \emph{not} diagonal in the Lorentz indices.

The self-energy renormalization is performed as described by Morningstar in
\cite{morning1}. The vertex corrections follow in a similar manner.
We will not write down all the Feynman rules here, but rather
will do a particular example: the one-loop amputated
vertex correction of the axial vector current at lowest order in $v^2$
in lattice NRQCD for the ``free \bc " system. In the notation of Eq.~(\ref{eq:lnrqcd}), this is 
$g^2\delta V_{lat}$.
This is a standard vertex correction: the axial vector current creates
a $c$ and $\overline{b}$ which then exchange a gluon between them. The one-loop amputated
vertex correction of the axial vector current at lowest order in $v^2$
in lattice NRQCD for the ``free \bc " system is
\be
g^2 \delta V_{lat}=\frac{4}{3}g^2\int_{-\pi}^\pi\frac{d^4 k}{(2\pi)^4}
\frac{D_{\mu\nu}(k) N_{\mu\nu}}
{\Delta_c(-k)\Delta_b(k)}
\;;\label{eq:vlatappendix}
\ee
the inverse propagator of the $c$ is ($F_c$ and $E_c$ are defined below; $n_c$ is a stability
parameter)
\be
\Delta_c(-k)=1-\exp(i k_4) F_c^{2n_c}(k)E_c^2(k)
\;;
\ee
the inverse propagator of the $\overline{b}$ is ($F_b$ and $E_b$ are defined below;
$n_b$ is a stability
parameter)
\be
\Delta_b(k)=1-\exp(-i k_4) F_b^{2n_b}(k)E_b^2(k)\;;
\ee
$D_{\mu\nu}$ is the above gluon propagator;
and $N_{\mu\nu}$, a product of  one gluon emission and absorption vertices, will be written below. 
There are no external momenta in Eq.~(\ref{eq:vlatappendix}) because at lowest order in $v^2$
the three-momenta of the $c$ and $\overline{b}$ are set to zero. Also recall that the
mass has been subtracted from the energies as part of the definition of NRQCD.
In the above inverse quark propagators 
\be
F(k)=1-\frac{\hat{\bf k}^2}{4 m n}
\ee
and
\be
E(k)=1-\sum_{i=1}^3 \frac{\hat{k}_i^4}{48 m}+\frac{(\hat{\bf k}^2)^2}{32 n m^2}
\;,
\ee
for the particular  $c$ and $\overline{b}$ masses and stability parameters.
The definitions
\be
\hat{k}_i=2 \sin (k_i/2)
\ee
 and 
 \be
 \hat{{\bf k}}^2=4\sum_{i=1}^3
 \sin^2 (k_i/2)
 \ee
 have been made.
The final piece is $N_{\mu\nu}$.
It is defined as
\be
D_{\mu\nu} N_{\mu\nu} = N_T+N_S+N_{ST}
\;,
\ee
where the `S' and `T' subscripts signify `spatial' and `temporal' 
components respectively.\footnote{
Recall that our gluon propagator is not diagonal in Lorentz indices---hence the spatial-temporal
terms.} These three terms are
\bea
&(1)&~~N_T=D_{44}(k) E_b(k)E_c(k)F_b^{n_b}(k)F_c^{n_c}(k)
\;,\nn\\
&(2)&~~N_S=\sum_{i,j=1}^3 D_{ij}(k) N_{ij}
\;,\nn\\
{\rm and}~~&(3)&~~N_{ST}=\sum_{i=1}^3 D_{i4}(k)\left(N_{i4}+N_{4i}\right)
\;,
\eea
where 
\bea
N_{ij}=\sin(k_i/2) \sin(k_j/2)&&\left[\raisebox{0ex}[3ex][2ex]{}
\frac{1}{2 m_c n_c} E_c(k) S_{n_c}^c(0,k)\left\{
1+\exp(i k_4) F_c^{n_c}(k)\right\}\right.\nn\\
&&\hspace{3ex}\left.+2 K_i^c(0,k)\left\{
1+\exp(i k_4) F_c^{2 n_c}(k)E_c(k)\right\}\raisebox{0ex}[3ex][2ex]{}\right]\nn\\
\times
&&\left[\raisebox{0ex}[3ex][2ex]{}\frac{1}{2 m_b n_b} E_b(k) S_{n_b}^b(0,k)\left\{
1+\exp(-i k_4) F_b^{n_b}(k)\right\}\right.\nn\\
&&\hspace{3ex}\left.+2 K_j^b(0,k)\left\{
1+\exp(-i k_4) F_b^{2 n_b}(k)E_b(k)\right\}\raisebox{0ex}[3ex][2ex]{}\right]
\;,
\eea
\be
S_n(p,k)=\sum_{\alpha=1}^n F^{\alpha-1}(p) F^{n-\alpha}(k)
\;,
\ee
\be
K_i(p,k)=\frac{1}{48 m}\left(\hat{p}_i^2+\hat{k}_i^2\right)
-\frac{1}{32 n m^2}\left(\hat{\bf p}^2+\hat{\bf k}^2\right)
\;,
\ee
\bea
N_{i4}=&&\left[\raisebox{0ex}[3ex][2ex]{}\frac{1}{2 m_c n_c} \sin(k_i/2) E_c(k) S_{n_c}^c(0,k)\left\{
1+\exp(i k_4) F_c^{n_c}(k)\right\}\right.\nn\\
&&\hspace{3ex}\left.+2 \sin(k_i/2) K_i^c(0,k)\left\{
1+\exp(i k_4) F_c^{2 n_c}(k)E_c(k)\right\}\raisebox{0ex}[3ex][2ex]{}\right]\nn\\
\times&&\left[\raisebox{0ex}[3ex][2ex]{}i\exp(-ik_4/2)E_b(k)F_b^{n_b}(k)\right]
\;,
\eea
and
\be
N_{4i}=\left.N_{i4}^\ast\right|_{b\leftrightarrow c}
\;.
\ee
The specification of the integrand of Eq.~(\ref{eq:vlatappendix}) is now complete.
\pagebreak
\section{one-loop axial vector correction}
\label{sec:appendixB}

This Appendix analyzes the origin of the `$1/v$' terms 
in Eq.~(\ref{eq:conttot}). First the NRQCD correction in Eq.~(\ref{e1}) is discussed,
and then we move on to the QCD correction in  Eq.~(\ref{e2}).

The NRQCD correction in Eq.~(\ref{e1}):
The integral of interest here is the
complete 
one-loop amputated axial vector correction of continuum NRQCD. 
The integral is straightforward and 
so we will not go into details here.
Its integration is clearly described around
Eqs.~(14)--(16) of {\footnotesize BF}.\footnote{In this Appendix we will
take ``Ref.~\cite{braaten}" $\longrightarrow$ ``{\footnotesize BF}".}
Note that the result is entirely
`$1/v$' terms [compare with Eq.~(\ref{ntnt})].

Now we discuss the correction in Eq.~(\ref{e2}).
The integral of interest is the
{\em infrared divergent piece} of the 
one-loop amputated axial vector correction in continuum QCD
(the other pieces are not discussed here because their integration is
straightforward).
The complete amputated vertex correction is written in Eq.~(11) of 
{\footnotesize BF}.\footnote{{\footnotesize BF}'s notation is used in this Appendix. To convert to
our notation take $\Lambda\longrightarrow g^2\del V$ and $\alpha_s\longrightarrow g^2/(4\pi)$.}
 The term of interest
is the first part  of the first term in this equation:
\be
\Lambda_{IR}=\frac{64\pi i \alpha_s}{3}\int_0^1 dx \int_0^{1-x} dy \,\mu^{2\eps}
\int\frac{d^Dk}{(2\pi)^D}\frac{2\,p\cdot p^\prime}{[k^2-(xp^\prime-yp)^2+i\veps]^3}
\;.\label{ir}
\ee 
The 4-momentum of the ${\overline b}$ is $p=\left(\sqrt{{\bf p}^2+m_b^2},{\bf p}\right)$, and in
the center-of-mass frame the 4-momentum of the $c$ 
is $p^\prime=\left(\sqrt{{\bf p}^2+m_c^2},-{\bf p}\right)$. After integrating over
$k$,  changing variables to $s=x+y$ and $t=x/s$, and then
integrating over $s$ (as described in {\footnotesize BF}), the result is
\be
\Lambda_{IR}=-\frac{2\alpha_s}{3\pi}m_b m_c \int_0^1 dt \frac{1}{\Delta_t-i\veps}\left[
\frac{1}{\eps_{IR}}+\log\left(\frac{{\tilde{\mu}}^2}{\Delta_t-i\veps}\right)\right]\left[
1+{\cal O}(v^2)\right]+{\cal O}(\eps_{IR})
\;,\label{bb}
\ee
where the usual $D=4-2\eps_{IR}$ replacement has been made;  for the $s$-integral to converge,
$\eps_{IR}$ has to
 be negative; 
 ${\tilde{\mu}}^2=4\pi\mu^2/e^\gamma$; and  $\Delta_t$ is given by
 \be
 \Delta_t=q^2 (t-t_+)(t-t_-)
 \;,
 \ee
 where
 \bea
 &&q^2=(p+p^\prime)^2=(m_b+m_c)^2+m_b m_c v^2+{\cal O}(v^4)\\
 {\rm and}~~~~~&&t_{\pm}=\frac{1}{q^2}\left[m_b^2+p\cdot p^\prime \pm 
 \sqrt{(p\cdot p^\prime)^2-m_b^2 m_c^2}\,\right]\nn\\
 &&~~~~=\frac{m_b\pm m_{red} v}{m_b+m_c}+{\cal O}(v^2)
 \;.\label{bbk}
 \eea
 Recall $1/m_{red}=1/m_b+1/m_c$. The origin of the `$1/v$' terms is seen by
 rewriting
 \be
 \frac{m_bm_c}{\Delta_t-i\veps}=\frac{m_bm_c}{q^2(t_+-t_-)}\left(
 \frac{1}{t-t_+-i\veps}-\frac{1}{t-t_-+i\veps}
 \right)
 \;,
 \ee
 and noting that
 \be
 \frac{m_bm_c}{q^2(t_+-t_-)}=\frac{1}{2v}+{\cal O}(v)
 \;;
 \ee
Note that the ${\cal O}(v^2)$ terms of Eq.~(\ref{bbk}) cancel in the difference: 
\be
t_+-t_-=\frac{2 m_{red}v}{m_b+m_c}+{\cal O}(v^3)
\;.
\ee
 
To proceed with the integration,
we find Lewin's book on polylogarithms \cite{lewin} helpful.
More specifically, we use the following three equations:
\begin{mathletters}
\label{generallabel}
\bea
\int_0^u\frac{\log(a+b t)}{c+e t}\,dt&=&
\frac{1}{e}\, \log\left(\frac{a e-b c}{e}\right) \log\left(\frac{c+e u}{c}\right)
-\frac{1}{e} \,{\rm Li_2}\left[\frac{b (c+e u)}{b c-a e}\right]\nn\\
&&\hspace{1in}\mbox{}
+\frac{1}{e}\,{\rm Li_2}\left(\frac{b c}{b c-a e}\right)\;,\; bc-ae\neq 0
\;,
\eea
\be
{\rm Li_2}(x\pm i \veps) \stackrel{x>1}{=}\frac{\pi^2}{3}-\frac{1}{2}\,\log^2(x)\pm i \pi 
\log(x)
-\left[\frac{\left(\frac{1}{x}\right)}{1^2}+\frac{\left(\frac{1}{x}\right)^2}{2^2}+
\frac{\left(\frac{1}{x}\right)^3}{3^2}+\cdots\right]
\;,
\ee
and
\be
{\rm Li_2}(x) \stackrel{x<-1}{=}-\frac{\pi^2}{6}-\frac{1}{2}\log^2(-x)
-\left[-\frac{\left(\frac{1}{-x}\right)}{1^2}+\frac{\left(\frac{1}{-x}\right)^2}{2^2}
-\frac{\left(\frac{1}{-x}\right)^3}{3^2}+\cdots\right]
\;,
\ee
\end{mathletters}
 where
$a$, $b$, $c$, and $e$ may be complex, but $x$ is real.
  ${\rm Li_2}(z)$ is the dilogarithm function nicely elucidated in
Ref.~\cite{lewin}.
  
  The integration of Eq.~(\ref{bb}) is now
  straightforward with result (as $v\rightarrow 0$)
  \bea
  \Lambda_{IR}&=&\frac{2\alpha_s}{3\pi} \left[\raisebox{0ex}[3ex][2ex]{}\frac{1}{\eps_{IR}}-2
  +2\log\frac{\tilde{\mu}}{m_b+m_c}
-2\,\frac{m_{red}}{m_b}\log\frac{m_{red}}{m_c}
-2\,\frac{m_{red}}{m_c}\log\frac{m_{red}}{m_b}
\right.\nn\\
&&~~~~~\left.\mbox{}
+\frac{\pi^2}{v}-\frac{i\pi}{v}\left(
\frac{1}{\epsilon_{IR}}-2\,\log\frac{2 m_{red} v}{\tilde{\mu}}
\right)\raisebox{0ex}[3ex][2ex]{}
\right]
  \;,\label{aaa}
  \eea
  where recall $\log({\tilde{\mu}}^2)=\log(\mu^2)+\log(4\pi)-\gamma$.
  Note that all the `$1/v$' terms of the full answer [Eq.~(\ref{e2})] are here,
  as was to be shown. The \emph{imaginary} `$1/\eps_{IR}$' term is also here,
  but there appears to be an extra \emph{real}  `$1/\eps_{IR}$' term.
  The fact that this is not an ``extra term" is seen after including the wave 
  function renormalization [Eq.~(10) of {\footnotesize BF}]:
  \be
  \sqrt{Z_Q}=1+\frac{2\alpha_s}{3\pi}\left[
  -\frac{1}{4\eps_{UV}}-\frac{1}{2\eps_{IR}}
  +\frac{3}{2}\log\frac{m_Q}{\tilde{\mu}}-1
  \right]
  \;;
  \ee
  both the ${\overline b}$ and $c$ contribute one of these factors which results in the
  cancelation of the \emph{real}  `$1/\eps_{IR}$' term of Eq.~(\ref{aaa}). 
  After including both of these wave function renormalization factors and the complete
  amputated vertex correction, the result is Eq.~(\ref{e2}) as mentioned in the body of the paper.


\newpage

\newpage

\newpage

\newpage

\newpage

\begin{table}
\caption{Comparison of  perturbative (one loop) and nonperturbative Landau links using
the boosted definition of the coupling [see Eq.\ (\ref{eq:bcoup})]. 
Recall $u_0=1-\alpha \,u_0^{(2)}$. 
\label{tab:u0}}
\begin{tabular}{ccccc}
 &&&\multicolumn{2}{c}{$u_0^{(2)}$}\\
 \cline{4-5}
 $\beta$
&$u_0=\left\langle\frac{1}{3} Re Tr U_\mu\right\rangle$&
$\alpha=10/(4\pi\beta u_0^4)$&nonpert&pert\\ \tableline
7.2&.805&.2632&.7409&.7503\\
7.3&.817&.2447&.7480&.7503\\
7.4&.8286&.2281&.7513&.7503\\
 \end{tabular}
 \end{table}

\begin{table}
\caption{One-loop perturbative lattice NRQCD matching factors of quarkonia. 
 $Z_{\rm match}$ is defined in and above Eq.\ (\ref{eq:heymatch}).
The results of the middle column
are to be multiplied by $4/[3(2\pi)^4]$. The statistical error estimate from {\footnotesize   VEGAS}
 is less than one percent.
\label{tab:zmatchquarkonia}}
\begin{tabular}{ccc}
ma[n]&$\del \overline{V}_{lat}+\del {Z}_{lat}^Q$&$Z_{\rm match}$\\ \tableline
1.4[3]&-43.9335&0.9141\\
1.5[3]&-42.3631&0.9037\\
1.6[3]&-45.3303&0.9049\\
2.1[2]&-49.3386&0.9274\\
2.2[2]&-49.8716&0.9235\\
2.3[2]&-50.3880&0.9192\\
2.8[2]&-52.8248&0.9359\\
2.9[2]&-53.2263&0.9323\\
3.0[2]&-53.7950&0.9288\\
3.5[2]&-57.0103&0.9462\\
3.6[2]&-57.5486&0.9437\\
3.7[2]&-58.3099&0.9416\\
4.2[2]&-61.5731&0.9574\\
4.3[2]&-62.3861&0.9564\\
4.4[2]&-63.3444&0.9558\\
 \end{tabular}
 \end{table}
 
 \begin{table}
\caption{One-loop perturbative lattice NRQCD matching factors of the \bc . 
 $Z_{\rm match}$ is defined in and above Eq.\ (\ref{eq:heymatch}).
The results of the middle column
 are to be multiplied by $4/[3(2\pi)^4]$. The statistical error estimate from {\footnotesize   VEGAS}
 is less than one percent.
\label{tab:zmatchbc}}
\begin{tabular}{ccc}
${\rm m_c}a[n],\,{\rm m_b}a[n]$&$\del \overline{V}_{lat}+\del {Z}_{lat}^c/2+
 \del{Z}_{lat}^b/2$&$Z_{\rm match}$\\ \tableline
1.4[3],\,3.8[2]&-47.5372&1.0048\\
1.5[3],\,4.4[2]&-47.2181&1.0096\\
1.6[3],\,5.0[2]&-49.3620&1.0213\\
 \end{tabular}
 \end{table}
 
\begin{table}
\caption{$M_{\rm pert}$ versus $M_{\rm kin}$ for runs tuned to 6.35 GeV for the
spin averaged
\bc . The statistical error estimate of $M_{\rm pert}$ from 
{\footnotesize VEGAS} is less than one percent.\label{tab:bc}}
\begin{tabular}{ccccccc}
 &&&&\multicolumn{2}{c}{$M_{\rm pert}\,a$}&\\
 \cline{5-6}
 $\beta$&a(fm)&${\rm m_c}a[n],\,{\rm m_b}a[n]$&$M_{\rm kin}\,a$
&w/o tad imp&w tad imp&$M_{\rm kin}$(GeV)\\ \tableline
$7.2$   &	$.204(1)$	& $1.6 [3],\,5.0[2] $ & $6.60(21)$ & $6.91 $ & $6.92 $ &$6.37(21)$\\
$7.3$   &  	$.185(2)$	& $1.5 [3],\,4.4[2] $ & $5.94(16)$ & $6.15$ & $6.28$ &$6.32(18)$\\
$7.4$   &    $.163(3)$	& $1.4 [3],\,3.8[2] $ & $5.26(18)$ & $5.42 $ & $5.67 $ &$6.37(24)$\\
 \end{tabular}
 \end{table}

\begin{table}
\caption{$M_{\rm pert}$ versus $M_{\rm kin}$ for  runs nearest the
charm and bottom regions respectively. The statistical error estimate 
of $M_{\rm pert}$ from {\footnotesize   VEGAS} is less than one percent.\label{tab:candb}}
\begin{tabular}{ccccccc}
 &&&&\multicolumn{2}{c}{$M_{\rm pert}\,a$}&\\
 \cline{5-6}
 $\beta$&a(fm)&ma[n]&$M_{\rm kin}\,a$
&w/o tad imp&w tad imp&$M_{\rm kin}$(GeV)\\ \tableline
$7.2$   &	$.240(3)$	& $1.6 [3] $ & $3.37(6)$ & $3.17$ & $3.89$ &$2.77(6)$\\
$7.3$   &  	$.205(3)$	& $1.5 [3]$ & $3.20(5)$ & $2.96$ & $3.69$ &$3.08(6)$\\
$7.4$   &    $.178(3)$	& $1.4 [3]$ & $3.00(6) $ & $2.80$ & $3.55$ &$3.32(9)$\\
$7.2$ &	$.198(2)$	& $4.4 [2]$ & $9.07(21)$ & $9.27$ & $8.78$ &$9.05(23)$\\
$7.3$   &    $.174(3)$	& $4.3 [2]$ & $8.74(20)$ & $9.06$ & $8.63$ &$9.91(28)$\\
$7.4$ &	$.151(2)$	& $3.5 [2]$ & $7.05(20)$ & $7.35$ & $7.19$ &$9.20(29)$\\
 \end{tabular}
 \end{table}
 
 \begin{table}
\caption{Vector decay constants of quarkonia (after including
perturbative matching).\label{tab:quarkoniadecay}}
\begin{tabular}{ccccc}
 &&&\multicolumn{2}{c}{$f_{V}$ (MeV)}\\
 \cline{4-5}
 $\beta$&ma[n], $M_{\rm kin}\,a$
&a(fm)&w/o $v^2$ rel cor&w $v^2$ rel cor\\ \tableline
7.4&1.4[3], 3.00(6)&.178(3)&411(8)&364(9)\\
7.3&1.5[3], 3.20(5)&.205(3)&396(7)&352(8)\\
7.2&1.6[3], 3.37(6)&.240(3)&375(6)&335(8)\\
7.4&2.1[2], 4.33(10)&.164(2)&492(8)&464(10)\\
7.3&2.2[2], 4.50(7)&.188(2)&474(6)&446(8)\\
7.2&2.3[2], 4.69(8)&.220(2)&440(6)&415(8)\\
7.4&2.8[2], 5.66(15)&.156(2)&561(10)&541(12)\\
7.3&2.9[2], 5.84(11)&.179(2)&533(8)&515(10)\\
7.2&3.0[2], 6.06(13)&.208(5)&495(13)&477(14)\\
7.4&3.5[2], 7.05(20)&.151(2)&619(12)&605(14)\\
7.3&3.6[2], 7.26(15)&.174(2)&581(9)&567(11)\\
7.2&3.7[2], 7.52(17)&.201(2)&538(8)&524(10)\\
7.4&4.2[2], 8.51(29)&.149(2)&662(14)&651(15)\\
7.3&4.3[2], 8.74(20)&.174(3)&608(12)&597(13)\\
7.2&4.4[2], 9.07(21)&.198(2)&567(9)&557(11)\\
\end{tabular}
 \end{table}
 
\begin{table}
\caption{\bc\ decay constants (after including
perturbative matching).\label{tab:bcdecay}}
\begin{tabular}{cccc}
 &&\multicolumn{2}{c}{$f_{B_c}$ (MeV)}\\
 \cline{3-4}
 $\beta$
&a(fm)&w/o $v^2$ rel cor&w $v^2$ rel cor\\ \tableline
7.2&.204(1)&462(8)&407(11)\\
7.3&.185(2)&463(8)&406(10)\\
7.4&.163(3)&478(12)&420(13)\\
 \end{tabular}
 \end{table}

\end{document}